\documentclass[graybox]{svmult}

\usepackage{mathptmx}       
\usepackage{helvet}         
\usepackage{courier}        
\usepackage{type1cm}        
\usepackage{graphics}
\usepackage{makeidx}         
\usepackage{graphicx}        
\usepackage{multicol}        
\usepackage[bottom]{footmisc}

\usepackage{subcaption}
\usepackage{caption}

\usepackage{amsmath}
\usepackage{amssymb}
\usepackage{graphicx}
\usepackage{graphicx, wrapfig, setspace, booktabs}

\makeindex             

\begin{document}

\title*{Group pattern detection of longitudinal data using   functional statistics} 
\author{Rongjiao Ji, Alessandra Micheletti, Natasa Krklec Jerinkic, Zoranka Desnica}
\institute{Rongjiao Ji, Alessandra Micheletti \at Universit\'a degli Studi di Milano, \email{rongjiao.ji@unimi.it, alessandra.micheletti@unimi.it}
\and Natasa Krklec Jerinkic \at University of Novi Sad,  \email{natasa.krklec@dmi.uns.ac.rs}
\and Zoranka Desnica \at 3Lateral DOO, \email{zoranka.desnica@3lateral.com} }
%
%

\maketitle

\abstract{Estimations and evaluations of the main patterns of time series data in groups benefit large amounts of applications in various fields. Different from the classical auto-correlation time series analysis and the modern neural networks techniques, in this paper we propose a combination of functional analysis of variance (FANOVA) and permutation tests in a more intuitive manner for a limited sample size. First, FANOVA is applied in order to separate the common information and to dig out the additional categorical influence through paired group comparison, the results of which are secondly analyzed through permutation tests to identify the time zones where the means of the different groups differ significantly. Normalized kernel functions of different groups are able to reflect remarkable mean characteristics in grouped unities, also meaningful for deeper interpretation and group-wise classification.
In order to learn whether and when the proposed method of FANOVA and permutation F-test works precisely and efficiently, we compare the estimated kernel results with the ground truth on simulated data. 
After the confirmation of the model's efficiency from simulation, we apply it also to the RAVDESS facial dataset to extract the emotional behaviors of humans based on facial muscles contractions (so-called action units (AU) technically in computer graphics), by comparing the neutral performances with emotional ones. 
} 




\section{Introduction}
Many natural, economical, or technological phenomena can be represented as real functions $f:\mathbb{R}\to \mathbb{R}$ where the domain can be time or another one-dimensional variable, and data of many copies of these functions can be collected, usually in a discrete format.

We here consider the case where we collect data of a set of random functions describing a real phenomenon, which is characterized by the presence of groups, where each group is characterized by a different mean underlying pattern that we want to detect.

Examples of such situations are: the collection of temperature evolution in different regions of the world, in the same time window, where we want to study the underlying temperature evolution functions characterizing each region and their main differences; the collection of stock price evolution, with stocks related with goods of different economical sectors, or different segments of the same sector, and we want to extract and compare the underlying evolution patterns of the sector or segments; etc.

In this paper we will face the problem of underlying pattern detection and comparison of groups of random functions, using a specific driving example: the detection of human emotions from functions describing the movements of facial muscles during a video. 

The study of human facial expressions and emotions never stops in our daily life while we communicate with others. Following the increased interest in automatic facial behavior analysis and understanding, the need of a semantic interpretation of the evolution of facial expressions and of human emotions has become of interest in recent years \cite{Fridlun2014}. In this paper, based on a work cooperated with the Serbian company 3Lateral, which has special expertise on building visual styles and designs in animation movies, we want to explore functional statistical instruments to identify the emotions while analyzing the expressions through recorded videos of human faces. The final aim of this research is to use this information to better and more realistically establish virtual digital characters, able to interact autonomously with real humans. 

The data that we consider are multivariate longitudinal data, showing the evolution in time of different face muscles contraction. Functional Data Analysis (FDA) offers the possibility to analyze the entire expression evolution process over time and to gain detailed and in-depth insight into the analysis of emotion patterns. The basic idea in functional data analysis is that the measured data are noisy observations coming from a smooth function. Ramsay and Silverman \cite{ramsay1997functional} describe the main features of FDA, that can be used to perform exploratory, confirmatory or predictive data analysis. Ullah and Finch \cite{UllahShahid2013} published a systematic review on the applications of functional data analysis, where they included all areas where FDA was applied.

Here we will apply models of functional regression and functional analysis of variance (FANOVA) to identify the  mean functional patterns related with different emotions. Additionally we setup a technique, based on permutation tests, to identify the time zones where the identified patterns are significantly different. The identified patterns can then be used to perform an automatic detection and classification of the emotions expressed by a human face.

The methods here proposed are also tested on simulated data to highlight their effectiveness in identifying the right underlying patterns and the correct time zones in which the patterns differ, also in relationship with the level of noise present in the collected data.

In spite of the specific motivating case study here considered, the methods here proposed and the related software can be applied to many other similar situations in different fields of application.

\section{Methods}



\subsection{Functional Data Analysis}

\textbf{Functional Data Setting} 
Let $\Omega$ be a population space of statistical units ($\omega \in \Omega$) and $s_n = \{\omega_1, \omega_2,\dots, \omega_n\}$ be a random sample of size $n$ drawn from $\Omega$. Let $\mathbf{Y}$ be a $L_2$-continuous multivariate stochastic process defined on $\Omega$ associating to a set of $D$ curves, $D \geq 2$, where each curve in one dimension defined on a given finite interval $\mathcal{T} = [0, T], 0 < T < \infty$, i.e.,
$\mathbf{Y}(\omega) = [\mathbf{y}_{\omega,1}(t), \dots, \mathbf{y}_{\omega,D}(t)]^T$ (denoted as $\mathbf{Y} = (\mathbf{y}_1,\dots, \mathbf{y}_D)^T$ for simplicity), $t \in \mathcal{T}$, such that
$$\lim\limits_{h\to 0} \mathbb{E} [||\mathbf{Y}(t+h) - \mathbf{Y}(t)||^2] = \lim\limits_{h\to 0}\int_0^T\sum\limits_{d=1}^D \mathbb{E}  [(\mathbf{y}_d(t+h) - \mathbf{y}_d(t))^2] dt = 0.$$
The observations of $\mathbf{Y}$ on the sample $s_n$ are called multivariate functional data, providing independent realizations as samples $\{\mathbf{Y}(\omega_1),\dots, \mathbf{Y}(\omega_n)\}$ with multivariate curves \cite{jacques2012clustering}. 
The mean functions of $\mathbf{y}_d$ and $\mathbf{Y}$ are
\begin{equation*}
    \mathbb{E}[\mathbf{y}_d(t)] = \mu_d(t), t \in [0, T], \text{ and } \mathbb{E[\mathbf{Y}]} = \mathbf{\mu}(t) = (\mu_1(t), \dots,\mu_d(t), \dots,\mu_D(t))^T .
\end{equation*}
The covariance operator of $\mathbf{Y}$ is then defined as an integral operator $\mathcal{C}$ with kernel
$$\mathbf{C}(t,s) = \mathbb{E}[(\mathbf{Y}(t)-\mathbf{\mu}(t))\otimes (\mathbf{Y}(s)-\mathbf{\mu}(s))], \forall t,s \in [0, T], $$
where $\otimes$ is the tensor product on $\mathbb{R}^D$. Thus, $\mathbf{C}(t,s)$ is a $D \times D$ matrix with elements
$$\mathbf{C}(t,s)[i,j] = Cov(\mathbf{Y}_i(t),\mathbf{Y}_j(s)), i,j = 1, \dots,D.$$


\textbf{Functional Principal Component Analysis (FPCA)} 
\label{sec:methoed - alignment}
Under the hypothesis of $L_2$-continuity, $\mathcal{C}$ is an Hilbert-Schmidt operator, i.e., it is compact and self-adjoint. Meanwhile, the spectral analysis of $C$ provides a countable set of positive eigenvalues $\{\eta_{j, j \geq 1}\}$ associated to an orthonormal basis of multivariate eigen-functions $\{\mathbf{f}_{j,j\geq 1}\}$, $\mathbf{f}_j = [\mathbf{f}_{j,1},\dots, \mathbf{f}_{j,D}]^T$, such that
$$C\mathbf{f}_j = \eta_j\mathbf{f}_j,$$
with $\eta_1 \geq \eta_2 \geq \dots$ and $\langle \mathbf{f}_i,\mathbf{f}_j \rangle _{\{L_2([0,T])\}^D} = \delta_{i,j}$ with $\delta_{i,j} = 1$ if $i=j$, otherwise $\delta_{i,j} =0$  \cite{berrendero2011principal}. Assume that the principal components $\eta_j$ of $\mathbf{Y}$ are zero-mean random variables, as the projections of $\mathbf{Y}$ on the $j$th eigenfunction of $\mathbf{C}$:

\begin{equation}
    \eta_j = \int_0^T \langle \mathbf{Y}(t) - \mathbf{\mu}(t),  \mathbf{f}_j (t) \rangle dt = \int_0^T \sum\limits_{d=1}^D(Y_d(t) - \mu_d(t))\mathbf{f}_{j,d}(t)dt,
\end{equation}

which means $\mathbf{Y}(t)$ varies around the mean function with random amplitude variations in the directions of $\{ \mathbf{f}_j \}_{j \geq 1}$. Therefore, following Karhunen-Loeve expansion, it holds that

\begin{equation}
    \mathbf{Y}(t) = \mathbf{\mu}(t) + \sum\limits_{j \geq 1}\eta_j\mathbf{f}_j (t)
    \text{ and }
    \mathbf{y}_d(t) = \mu_d(t) + \sum\limits_{j \geq 1}\eta_j \mathbf{f}_{j,d} (t)
\end{equation}
are the best approximations of this form under the mean square criterion.


\textbf{Functional Registration}
In practice, it is essential to align observed functions into a common registered internal timeline to isolate the phase variability of the functions, but keeping, at the same time, the amplitude-phase unchanged to maintain the information under interests. The phase variation is normally represented by a random change of time scale, which is mostly a non-linear transformation. 
We use the warping functions $T_i: [0,T] \rightarrow [0,T], i = 1, \dots, n$, assuming that they are increasing functions independent of amplitude variation. They map unregistered chronological time $t_i^*$ to registered internal time $t$ so that $T_i^{-1}(t_i^*) = t$, with $E[T_i(t)] = t$. The observed time-warped curves, represented through a Karhunen-Loeve expansion based on a functional basis $\mathbf{f}_{j,d}$, are
$$\tilde{y}_{i,d}(t) = y_{i,d}(T_i^{-1}(t_i^*)) = \mu_{d}(T_i^{-1}(t_i^*)) + \sum\limits_{j \geq 1} \eta_j\mathbf{f}_{j,d} (T_i^{-1}(t_i^*)).$$
The literature is rich of proper methods to do this, and a more in-depth review would be beyond the scope of this report. Briefly speaking, we followed the principal components based registration method \cite{wrobel2019register} with a generative process, by alternating the process between using the generalized FPCA to calculate template functions and estimating smooth time warping functions that map observed curves to templates \footnote{Codes are available in the R package "registr" \cite{wrobel2018register}}.





\textbf{Approximation into finite basis of functions}
Often in practice, data are observed at discrete time points and with some noise. In order to get the functional feature of data, researchers normally use some smoothing and interpolation methods to rebuild the consecutive function from the discrete data, while assuming that the latent function belongs to a finite dimensional functional space spanned by some functional basis. In this way, functional data objects are constructed from time series data by specifying a set of basis functions and a set of coefficients defining a linear combination of these basis functions. For non-periodic time series, B-spline basis functions which contain polynomial segments jointed end-to-end at argument values are frequently considered, benefiting the most from fast computational ability and great flexibility \cite{ramsay1997functional}.

\subsection{Multiple Multivariate Function-on-Scalar Regression}

Let $y_{g,d,k}(t)$ denote the evolution of one specific variate $d \in \{1,\dots,D\}$ in sample $k \in \{1,\dots,K\}$ for one group $g\in \{0,\dots, G\}$. We can assume that
\begin{equation}
\label{flm:indiv0}
y_{g,d,k}(t) = \mu_{d,0}(t) + \alpha_{d,g}(t) + \epsilon_{g,d,k}(t),
\end{equation}
where $\mu_{d,0}(t)$ is the grand mean function independent from groups. The term $\alpha_{d,g}(t)$ is the additional effect function of group $g$ on the considered variate $d$, while $\epsilon_{g,d,k}(t)$ represents the unexplained zero-mean variation, specific of the $k$-th sample within group $g$. In order to identify the parameter functions uniquely, we require the constraint $\sum\limits_{g=0}^G \alpha_{d,g}(t) = 0, \forall t.$

By grouping the samples from the same group, we can define a $(GK+1) \times (G+1)$ design matrix $\mathbf{Z}_d$ for univariate model, with suitable 0 and 1 entries, as described in \cite[Section 9.2]{ramsay1997functional}, and rewrite Equation (\ref{flm:indiv0}) in matrix form
$$\mathbf{y}_d(t)=\mathbf{Z}_d\mathbf{\beta}_d(t) + \mathbf{\epsilon}_d(t), \forall t,$$ 
where
$$\mathbf{y}_d(t)=\begin{bmatrix}
y_{0,d,1}(t) \\
\vdots \\
y_{0,d,K}(t) \\
y_{1,d,1}(t) \\
\vdots \\
y_{1,d,K}(t) \\
\vdots \\
 y_{G,d,1}(t) \\
\vdots \\
y_{G,d,K}(t) \\
0
\end{bmatrix},
\mathbf{Z}_d=
\begin{bmatrix}

1 & 1 & 0 & 0 &   \cdots & 0 & 0 \\ 
\vdots & \vdots  & \vdots & \vdots & \cdots & \vdots & \vdots \\
1 & 1 & 0 & 0 &   \cdots & 0 & 0 \\ 
1 & 0 & 1 & 0 &   \cdots & 0 & 0 \\ 
\vdots & \vdots &   \vdots & \vdots & \cdots & \vdots & \vdots \\
1 & 0 & 1 & 0 &   \cdots & 0 & 0 \\ 
  &   &   &     & \vdots &   &  \\
1 & 0 & 0 & 0 &   \cdots & 0 & 1 \\ 
\vdots & \vdots   & \vdots & \vdots & \cdots & \vdots & \vdots \\
1 & 0 & 0 & 0 &   \cdots & 0 & 1 \\
0 & 1 & 1 & 1 &   \cdots & 1 & 1
\end{bmatrix},
\mathbf{\beta}_d(t) = \begin{bmatrix}
\mu_{d,0}(t) \\
\alpha_{d,0}(t)\\
\alpha_{d,1}(t)\\
\vdots \\
\alpha_{d,G}(t)
\end{bmatrix} = \begin{bmatrix}
\beta_{d,0}(t) \\
\beta_{d,1}(t) \\
\beta_{d,2}(t)\\
\vdots \\
\beta_{d,G+1}(t)
\end{bmatrix},
\mathbf{\epsilon}_d(t)=\begin{bmatrix}
\epsilon_{0,d,1}(t) \\
\vdots \\
\epsilon_{0,d,K}(t) \\
\epsilon_{1,d,1}(t) \\
\vdots \\
\epsilon_{1,d,K}(t) \\
\vdots \\
 \epsilon_{G,d,1}(t) \\
\vdots \\
\epsilon_{G,d,K}(t) \\
0
\end{bmatrix}.
$$

Here $\mathbf{\beta}_d$ is defined as the corresponding set of $G+1$ regression functions for mean and $G$ additional groups, and $\mathbf{\epsilon}_d$ is the matrix of noise with the same dimension of $\mathbf{y}_d$.

Following the notations above, we can write the multivariate matrix form for the multiple regression functions, as 

$$\mathbf{Y}(t) = \mathbf{Z}\mathbf{\beta}(t) + \mathbf{\epsilon}(t),$$
where
$$\mathbf{Y}(t) = \begin{bmatrix}
\mathbf{y}_1(t) \\
\vdots \\
\mathbf{y}_d(t) \\
\vdots \\
\mathbf{y}_D(t)
\end{bmatrix},
\mathbf{Z} = \begin{bmatrix}
\mathbf{Z}_1 \\
\vdots \\
\mathbf{Z}_d \\
\vdots \\
\mathbf{Z}_D
\end{bmatrix}^T,
\mathbf{\beta}(t) = \begin{bmatrix}
\mathbf{\beta}_1(t) \\
\vdots \\
\mathbf{\beta}_d(t) \\
\vdots \\
\mathbf{\beta}_D(t)
\end{bmatrix},
\mathbf{\epsilon}(t) = \begin{bmatrix}
\mathbf{\epsilon}_1(t) \\
\vdots \\
\mathbf{\epsilon}_d(t) \\
\vdots \\
\mathbf{\epsilon}_D(t)
\end{bmatrix}.
$$

Under the functional least squares fitting criterion on $R^{GDK \times T}$ space, we can estimate the parameters  
\begin{equation}
   \hat{\mathbf{\beta}}(t) = \hat{\mathbf{B}}\Theta(t),
\end{equation}
where $\Theta(t) = (\theta_{1}(t), \dots, \theta_{q}(t), \dots, \theta_{Q}(t))^T$ denotes a linear combination of basis functions with $ J_{\theta \theta}$ as the inner product of each pair of functional basis, and $\hat{\mathbf{B}}$ is the estimated coefficient matrix with 
$$vec(\mathbf{B}^T) = [( \mathbf{Z} \otimes J_{\theta \theta}^{\frac{1}{2}} )^T (\mathbf{Z} \otimes J_{\theta \theta}^{\frac{1}{2}} )]^{-1} (\mathbf{Z} \otimes J_{\theta \theta}^{\frac{1}{2}} )^T vec(J_{\theta \theta}^{\frac{1}{2}} \mathbf{A}^T).$$
Here,  $\mathbf{A}$ is the coefficient matrix of samples $\mathbf{Y}(t)$ with a fixed B-spline functional basis and more computation details in matrix format are shown in Appendix \ref{Appendix:chapter1_fda}.

\subsection{Functional ANOVA and F-type tests}

After establishing the model for functional evolution processes and estimating the additional parameter functions affected by group categories, next goal is to investigate whether and when group labels are significantly influencing the change of variables. ANOVA-related methods are suitable tools to determine the effects of independent categorical variables on continuous dependent variables, by comparing inter-group differences between centroids represented as the vectors of the mean values of the dependent variables. Moreover, in order to construct feasibly more general hypothesis tests for group means, contrasts, defined as a linear combination of regression coefficients where the values of the linear combination sum up to 0, are applied by specifying the values on different pairs of groups gradually.

For each subgroup of variable $d \in \{1,\dots,D\}$ and group $g \in \{0,\dots,G\}$, assume that the $k$-th sample curve in this subgroup $y_{d,g,k}(t) \sim \mathbb{N}(\mu_{d,g}(t),\sigma^2), k \in \{1,\dots, K\}$ with group mean $\mu_{d,g}(t)$ and constant variance $\sigma^2$, and therefore the subgroup sample mean $\overline{Y}_{d,g,\cdot}(t) = \frac{1}{K} \sum\limits_{k=1}^{K} y_{d,g,k}(t) \sim \mathbb{N}(\mu_{d,g}(t),\frac{\sigma^2}{K})$. Here denote $\mathbf{\mu}(t)$ as the sequence of subgroup means $\mu_{d,g}(t)$ for all $d \in \{1,\dots,D\}$ and $g \in \{0,\dots,G\}$. We construct one test for each pair of groups in comparison (one control group $g=0$ and another group under study $g=\tilde{g}$ for one variable $d$) and in $i$-th test ($i \in \{1,\dots,DG\}$) we design a vector $\mathbf{a} = [0,\dots,0,1,0,\dots, 0,-1,0,\dots,0]$ in length $D(G+2)$ (same as the length of $\mathbf{\beta}(t)$ in last section) with the entries pointed to the pair of groups in comparison being 1 and -1 respectively, i.e., $a_{d(G+2)+2} = 1$, $a_{d(G+2)+2+\tilde{g}} = -1$, and 0 in entries remained. We define the contrast in $i$-th test as $c = \mathbf{a}^T \mathbf{\mu}(t)= \alpha_{d,0}(t) - \alpha_{d,\tilde{g}}(t) = (\alpha_{d,0}(t)+ \mu_{d,0}) - (\alpha_{d,\tilde{g}}(t)+\mu_{d,0})$ for $DG$ different group comparisons, where the sum of weigths in $\mathbf{a}$ is zero, and then the difference of subgroup sample means $\overline{Y}_{d,0,\cdot}(t) - \overline{Y}_{d,\tilde{g},\cdot}(t)$ is an unbiased estimator of the contrast $c$. Following this setting, the null hypothesis at each time point $t$ for the control group and the group under study $\tilde{g}$ in variable $d$ is 
$$\mathcal{H}_0: \mathbf{a}^T\beta(t) = 0 \Leftrightarrow  \alpha_{d,0}(t) - \alpha_{d,\tilde{g}}(t) = 0,$$
and the related F statistics under null hypothesis is 
$$F_{d,\tilde{g}}(t) = \frac{(\overline{Y}_{d,0,\cdot}(t) - \overline{Y}_{d,\tilde{g},\cdot}(t))^2}{S_p^2 \frac{2}{K}},$$
where the sum of sample variances is $S_p^2 = \frac{1}{N-DG} \sum\limits_{g=1}^G \sum\limits_{d=1}^D \sum\limits_{k=1}^K (y_{d,k,g}(t) - \overline{y}_{d,g,\cdot}(t))^2$ with sample size $N$. When assuming the functional data $y_{d,k,g}(t)$ as realizations of Gaussian processes, for any given $t \in \mathcal{T}$, $F_{d,\tilde{g}}(t) \sim F_{1,N-DG}$, where $F_{1,N-DG}$ denotes $F$-distribution with $1$ and $N-DG$ degrees of freedom.


The statistic $F_{d,\tilde{g}}(t)$ should be compared with a critical value of $F_{1, N-DG}$ for any predetermined significant level $\alpha$. The critical value of $F_{1,N-DG}(\alpha)$ represents the $\alpha$-fraction of rejection area, where $F_{1,N-DG}(\alpha)$ denotes the upper $100\alpha$ percentile of $F_{1,N-DG}$. We write the $1 - \alpha$ quantile of the distribution of F statistic by $c_{1-\alpha}$, and given $c_{1-\alpha}$, we reject the null hypothesis when $F_{d,\tilde{g}}(t) > c_{1-\alpha}$, which indicates the dissimilarity of group means under comparison. First, we can test the statistic at all points of $\mathcal{T}$ using the same critical value from standard F-Distribution table with correlated degrees of freedom. Since complex assumptions and convoluted formula are not often satisfied in practice for most classic statistical tests, we consider also a nonparametric test, permutation test, which is light on assumptions, widely applicable and intuitive friendly. The permutation distribution is obtained by shuffling the labels within the paired groups in comparison and calculate the F statistic $F_{d,\tilde{g}}(t)$ many times to approximate to the statistic's distribution. More inference of permutation test can be seen in Appendix \ref{Appendix:chapter1_permutationtest}.

\subsection{Kernel-based Classification}

\begin{figure}[h]
    \centering
    \includegraphics[angle=270,width=\linewidth]{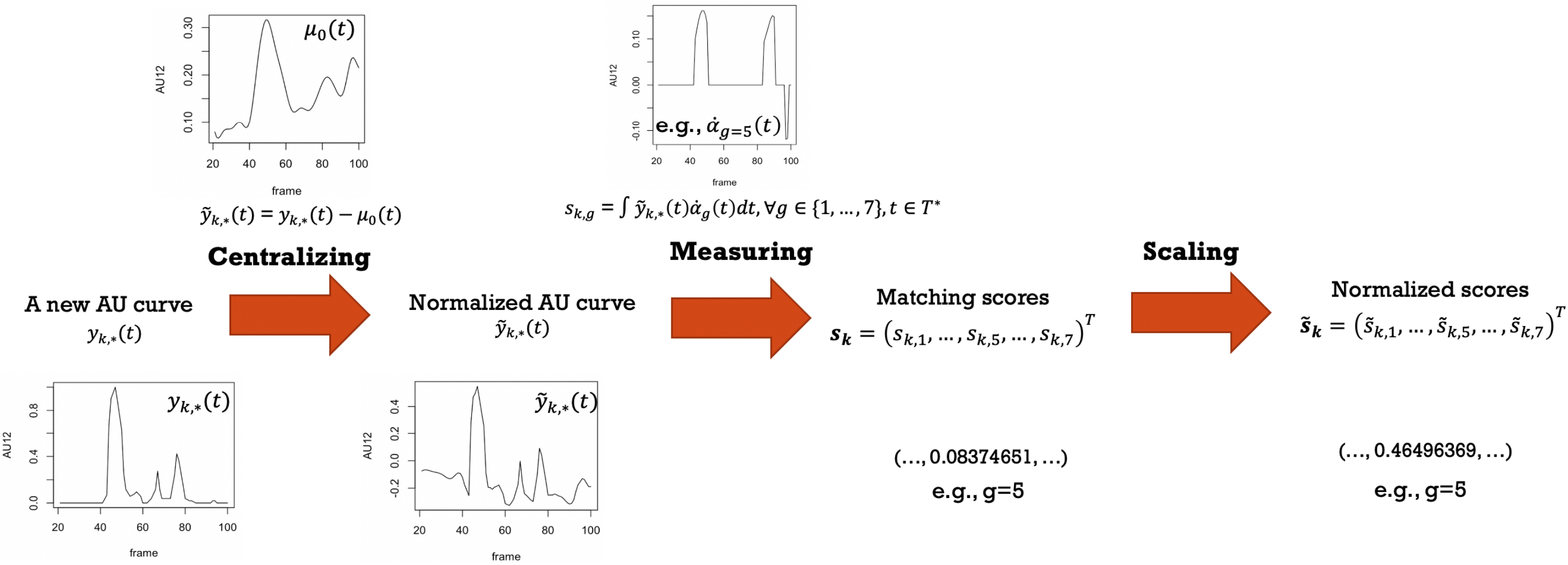}  
    \caption{Post hoc analysis by classifying from a given video}
    \label{fig:classification}
\end{figure}

After the explanation of functional linear regressions and test methods, we are able to obtain the ground mean functions and the additional group-wise functions revealing the latent group effects on variable curves. Most importantly, we can detect the time zones where the additional effects impact significantly in statistical sense. To further examine the efficiency of the information, a natural attempt would be to apply a post hoc analysis: do classifications using the knowledge obtained and check the correctness. Before the classification, four steps are needed in our framework to project the input functions into normalized scores in lower dimensions:

\begin{itemize}
    \item First, centralize the input evolution by subtracting the estimated mean function for each sample $k$ on variable $d$. 
    
    $$\tilde{y}_{d,k}(t) = y_{d,k}(t) - \mu_{d,0}(t)$$
    
    \item Secondly, use the centralized evolution to make inner product with $G$ estimated addtional group-wise functions (also called kernels for simplicity) in turn, and thus the evolution function is projected to a vector space with $G$ dimensions. Let $S_{g,d,k}$ represent the score of group $g$ on variable $d$ in sample $k$:
    
    $$S_{g,d,k} = (\tilde{y}_{d,k}(t), \hat{\alpha}_{g,d}(t)) = \int_0^T w_{g,d} \tilde{y}_{d,k}(t) \hat{\alpha}_{g,d}(t) dt = \int_I w_{g,d} \tilde{y}_{d,k}(t) \hat{\alpha}_{g,d}(t) dt,$$
    where $I$ represents the union of relevant intervals, and $w_{g,d}$ is the weights in the integral process, constant in time.

    \item Furthermore, normalize the scores across all groups and variables from one sample $k$, so that $\tilde{S}_{g,d,k} \in [0,1]$, in order to keep the relative rate/percentage across groups. 
    
    \begin{itemize}
        \item If $S_{g,d,k} < 0$ for some $g$, move them to positive numbers by subtracting the minimum ($ min\{0, min_g S_{g,d,k}\}$).
        \item Then, rescale them to be smaller than $1$ across groupss.
        $$\tilde{S}_{g,d,k} = \frac{ S_{g,d,k} - min\{0, min_g S_{g,d,k}\} }{ \sum\limits_{g=1}^G[S_{g,d,k} - min\{0, min_g S_{g,d,k} \}] }.$$
    \end{itemize}

    In this way $\tilde{S}_{g,d,k}$ represents the probability that variable $d$ in sample $k$ behaves as in (has been generated by) group $g$. If the sample $k$ that we are considering is labeled with $g$, more importance is given to the variables that assign a highest probability $\tilde{S}_{g,d,k}$ to the group $g$.

    \item Finally, combine the scores across variables for sample $k$ (labeled with $g$) by a linear convex combination with the trained weights $r_{l,g}$, such that, if the label of sample $k$ is $g$, then $S_{g,k}$ is the maximum score.

    $$S_{g,k}^{comb} = r_{1,g}\tilde{S}_{g,1,k} + r_{2,g}\tilde{S}_{g,2,k} + \dots + r_{L,g}\tilde{S}_{g,L,k}, $$
    with $r_{l,g} \in [0,1]$ and $\sum\limits_{l=1}^L r_{l,g} =1$. For sample $k$, $[S_{1,k}, S_{2,k}, \dots, S_{8,k}]$ contains the scores of all groups under study, and all the elements of this vector fall in region $[0,1]$.
    
\end{itemize}

For classification based on scores, we apply SVM method on scores $[S_{1,k}, S_{2,k}, \dots, S_{G,k}]$ to classify the samples and examine the kernels efficiency. A further study on the classification methods is important but is skippped here for space saving because the classification is not the focus of this paper, although we will explore further in next study.





\appendix

\section{Details of Methods}

\subsection{Approximation into a finite basis of functions}\label{Appendix:chapter1_fda}

Assume that each curve $y_{g,d,k} (d \in \{1, \dots, D\}$ for $k$-th video can be expressed as a linear combination of basis functions $\Theta(t) = (\theta_{1}(t), \dots, \theta_{q}(t), \dots, \theta_{Q}(t))^T$. We define $ J_{\theta \theta}$ to represent the inner product of each pair of functional basis. 

$$y_{g,d,k}(t)= \sum\limits_{q=1}^{Q} a_{g,d,k,q}\theta_{q}(t) =\mathbf{a}_{g,d,k}\Theta(t),$$

$$\mathbf{y}_{g,d}(t)= \mathbf{A}_{g,d} \Theta(t), \ \  \mathbf{A}_{g,d} = [\mathbf{a}_{g,d,1}, \dots, \mathbf{a}_{g,d,K}]^T$$

$$\mathbf{Y}_{g}(t)= \mathbf{A}_{g} \Theta(t), \ \  \mathbf{A}_{g} = [\mathbf{A}_{g,1}, \dots, \mathbf{A}_{g,d}, \dots, \mathbf{A}_{g,D}]^T$$

$$\mathbf{Y} = \mathbf{A} \Theta(t),\ \  \mathbf{A} = [\mathbf{A}_{1}, \dots, \mathbf{A}_{g}, \dots, \mathbf{A}_{G}]^T$$

$$\text{with }    \mathbf{A}_{g} =[\mathbf{A}_{g,1}, \dots,\mathbf{A}_{g,d}, \dots,\mathbf{A}_{g,D}]^T, 
\mathbf{A}_{g,d} = \begin{bmatrix}
    a_{g,d,1,1} & \dots & a_{g,d,1,Q}, \\
    \vdots & \vdots & \vdots \\
    a_{g,d,K,1} & \dots & a_{g,d,K,Q} \end{bmatrix}.$$

Similarly, the coefficient function $\mathbf{\beta} = [\beta_1(t), \dots, \beta_D(t)]^T$ can be decomposed with the same functional basis in the following way

$$\mathbf{\beta} = \mathbf{B} \Theta(t),$$
$$\text{with }    \mathbf{B} =[\mathbf{B}_1,\dots,\mathbf{B}_d, \dots,\mathbf{B}_D]^T, 
 \mathbf{B}_d = \begin{bmatrix}
    b_{1,d,1} & \dots & b_{1,d,Q}, \\
    \vdots & \vdots & \vdots  \\
    b_{G,d,1} & \dots & b_{G,d,Q} \end{bmatrix}.$$

Consider that in multivariate linear regression setting
$$\mathbf{Y} = \mathbf{Z} \mathbf{\beta} + \mathbf{E},$$
by substituting the functional curves with coefficient matrix and functional basis, we have

$$
\mathbf{A} \Theta(t) = \mathbf{Z} \mathbf{B} \Theta(t) + \mathbf{E}
$$

Notice that $\mathbf{B}$ is the coefficient that we want to estimate. Here by applying least square method on $R^{GDK \times T}$ space, we have %

\begin{align}
J(B) &= \| E \|_{L_2}^2 = \sum\limits_{\substack{g \in \{1,\dots,G\}, \\ d \in \{1,\dots,D\},\\  k \in \{1,\dots,K\}}} [y_{g,d,k}(t) - \hat{y}_{g,d,k}(t) ]_{L_2}^2 \\
     &= \int \| \mathbf{A} \Theta(t) -  \mathbf{Z} \mathbf{B} \Theta(t) \|_F^2 dt \\
     &= \int \| (\mathbf{A} - \mathbf{Z} \mathbf{B}) \Theta(t) \|_F^2 dt \\
     &= \int trace[((\mathbf{A} - \mathbf{Z} \mathbf{B}) \Theta(t))^T ((\mathbf{A} - \mathbf{Z} \mathbf{B}) \Theta(t))] dt \\
     &= \int trace[\Theta(t)^T(\mathbf{A} - \mathbf{Z} \mathbf{B})^T (\mathbf{A} - \mathbf{Z} \mathbf{B}) \Theta(t)] dt \\
     &= trace[ \int \Theta(t)^T(\mathbf{A} - \mathbf{Z} \mathbf{B})^T (\mathbf{A} - \mathbf{Z} \mathbf{B}) \Theta(t) dt] \\
     &= trace[ (\mathbf{A} - \mathbf{Z} \mathbf{B})^T (\mathbf{A} - \mathbf{Z} \mathbf{B}) \int \Theta(t)^T \Theta(t) dt] \\
     &= tr[( (\mathbf{A} - \mathbf{Z} \mathbf{B} ) J_{\theta \theta}^{\frac{1}{2}})^T (\mathbf{A} - \mathbf{Z} \mathbf{B} ) J_{\theta \theta}^{\frac{1}{2}} \\
     &= vec[( \mathbf{A} - \mathbf{Z} \mathbf{B} ) J_{\theta \theta}^{\frac{1}{2}}]^T  vec[( \mathbf{A} - \mathbf{Z} \mathbf{B} ) J_{\theta \theta}^{\frac{1}{2}}] \\
     &= \| vec[( \mathbf{A} - \mathbf{Z} \mathbf{B} ) J_{\theta \theta}^{\frac{1}{2}}]^T  \|_2^2 \\
     &= \| vec(J_{\theta \theta}^{\frac{1}{2}} \mathbf{A}^T) - vec[J_{\theta \theta}^{\frac{1}{2}} \mathbf{B}^T \mathbf{Z}^T ) \|^2 \\
     &= \| vec(J_{\theta \theta}^{\frac{1}{2}} \mathbf{A}^T) - (\mathbf{Z} \otimes J_{\theta \theta}^{\frac{1}{2}} ) vec(\mathbf{B}^T) \|^2, \\
\end{align}

Let $J(\mathbf{B}) = \mathbf{0}$, we can derive that

$$vec(\mathbf{B}^T) = [( \mathbf{Z} \otimes J_{\theta \theta}^{\frac{1}{2}} )^T (\mathbf{Z} \otimes J_{\theta \theta}^{\frac{1}{2}} )]^{-1} (\mathbf{Z} \otimes J_{\theta \theta}^{\frac{1}{2}} )^T vec(J_{\theta \theta}^{\frac{1}{2}} \mathbf{A}^T),$$
together with
\begin{align*}
    \mathbf{C}_{mat} 
    &= (\mathbf{Z} \otimes J_{\theta \theta}^{\frac{1}{2}} )^T (\mathbf{Z} \otimes J_{\theta \theta}^{\frac{1}{2}} ) \\
    &=  (\mathbf{Z}^T \mathbf{Z}) \otimes ({J_{\theta \theta}^{\frac{1}{2}}}^T J_{\theta \theta}^{\frac{1}{2}}) \\
    \mathbf{D}_{mat} &= (\mathbf{Z} \otimes J_{\theta \theta}^{\frac{1}{2}} )^T  vec(J_{\theta \theta}^{\frac{1}{2}} \mathbf{A}^T)
\end{align*}

\section{Permutation Test}\label{Appendix:chapter1_permutationtest}

Let's denote $\Phi_n$ as the set of all possible permutations of ${1,\dots,n}$, $M_n$ as the cardinality of $\Phi_n$. A permutation $\phi  \overset{\Delta}{=} (\phi_1,\dots,\phi_n) \in \Phi$.

$X_n^{\phi}$:  the permuted version of $X_n$, $X_n^{\phi} \overset{\Delta}{=} (X_1^{\phi},\dots, X_n^{\phi}) $.
$Fr_n^{\phi} = Fr_n(X_n^{\phi})$ to denote the F-test statistic computed based on the permutations $X_n^{\phi}$.  

Let $F_{Fr_n^{\phi}}(s) $ be the permutation distribution function of $Fr_n^{\phi}$ , defined as 
$$F_{Fr_n^{\phi}}(s) =  M_n^{-1} \sum\limits_{\phi \in \Phi_n} 1_{Fr_n(X_n^{\phi}) \leq s}. $$
 
We write the $1 - \alpha$ quantile of $F_{Fr_n^{\phi}}$  by $c_{1-\alpha,n}$, defined as $c_{1-\alpha,n} \overset{\Delta}{=} \inf \{t: F_{Fr_n^{\phi}}(s) \geq 1 -\alpha \} $.

Given the quantile $c_{1-\alpha,n}$, the permutation test rejects the null hypothesis when $Fr_n > c_{1-\alpha,n}$.



\section{Acknowledgements}
This work was funded by European Union’s Horizon 2020 research and innovation programme under the Marie Skłodowska Curie grant agreement No 812912 for the project BIGMATH. 
Computational resources have been provided by the INDACO Platform,
which is a project of High Performance Computing at the University of Milan.

\bibliographystyle{plain}
\bibliography{bibliography.bib}

\end{document}